\documentclass[aps,prl,twocolumn,superscriptaddress,showpacs,nofootinbib,longbibliography]{revtex4-1}
\usepackage{graphicx}
\usepackage{dcolumn}
\usepackage{bm}
\usepackage{hyperref}
\usepackage{upgreek}
\usepackage{color}
\usepackage{soul}
\usepackage{epstopdf}
\usepackage{units}
\usepackage{longtable}
\usepackage{floatrow}
\usepackage{latexsym}
\usepackage{gensymb}
\usepackage{floatrow}
\usepackage{mhchem}

\begin{document}

\title{Macroscopic Quantum Tunneling of a Topological Ferromagnet}

\author{Kajetan M. Fijalkowski}
\email[email:]{kajetan.fijalkowski@physik.uni-wuerzburg.de}
\author{Nan Liu}
\author{Pankaj Mandal}
\author{Steffen Schreyeck}
\author{Karl Brunner}
\author{Charles Gould}
\email[email:]{charles.gould@physik.uni-wuerzburg.de}
\author{Laurens W. Molenkamp}
\affiliation{Faculty for Physics and Astronomy (EP3), Universit\"at W\"urzburg, Am Hubland, D-97074, W\"urzburg, Germany}
\affiliation{Institute for Topological Insulators, Am Hubland, D-97074, W\"urzburg, Germany}

\date{\today}

\maketitle

\section[Abstract]{Abstract}

\textbf{The recent advent of topological states of matter spawned many significant discoveries. The quantum anomalous Hall effect\cite{Yu2010,Nomura2011,Chang2013} is a prime example due to its potential for applications in quantum metrology\cite{Goetz2018,Fox2018} as well as its influence on fundamental research into the underlying topological and magnetic states\cite{Lachman2015,Grauer2015,Liu2016,Yasuda2016,Fijalkowski2020,Fijalkowski2021b} and axion electrodynamics\cite{Nomura2011,tokura1,Grauer2017,Fijalkowski2021a}.
Here, we perform electronic transport studies on a (V,Bi,Sb)$_2$Te$_3$ ferromagnetic topological insulator nanostructure in the quantum anomalous Hall regime. This allows us access to the dynamics of an individual ferromagnetic domain. The volume of the domain is estimated to be about 85 000 nm$^3$, containing some 50 000 vanadium atoms, spread over a macroscopic distance of 115 nm. Telegraph noise resulting from the magnetization fluctuations of this domain is observed in the Hall signal. Careful analysis of the influence of temperature and external magnetic field on the domain switching statistics provides evidence for quantum tunneling of magnetization\cite{Chudnovsky1988,Awschalom1992,Coppinger1995,Friedman1996,Thomas1996,Sangregorio1997,Wernsdorfer1997,Barco1998} in a macrospin state. This ferromagnetic macrospin is not only the largest magnetic object in which quantum tunneling has been observed, but also the first observation of the effect in a topological state of matter.}

\begin{figure*}
\includegraphics[width=\columnwidth]{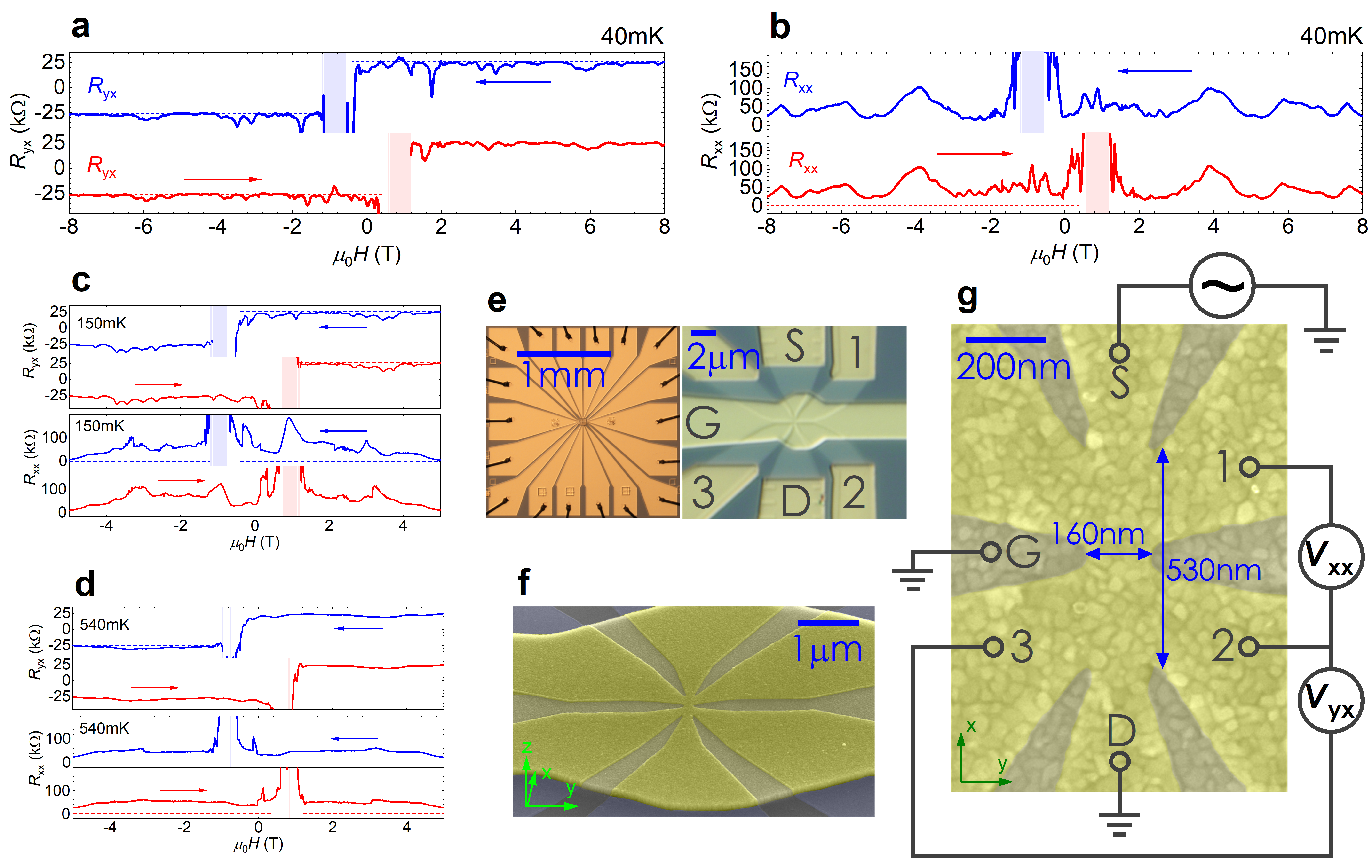}
\caption{
\textbf{Quantum anomalous Hall measurements on a (V,Bi,Sb)$_2$Te$_3$ nanostructure}. a) Hall resistance ($R_{\mathrm{yx}}$), and b) longitudinal resistance ($R_{\mathrm{xx}}$), collected at temperature 40 mK as a function of magnetic field. c-d) Same, for higher sample temperatures of 150 mK (c), and 540 mK (d). The  arrows indicate the magnetic field sweep direction. The shaded regions represent a regime around the global magnetization reversal where the two-terminal resistance of the sample exceeds 4 M$\Omega$, and the four-terminal measurement scheme becomes invalid. The horizontal dashed lines indicate the resistance values expected from a perfect quantum anomalous Hall effect. e) Optical microscope images of the actual device used in the experiment, left: a low magnification image of the entire sample, right: an image of the nanostructure with the contact labelling (S: source, D: drain, G: gate, and 1-3: voltage probes). f) False color scanning electron microscope (SEM) image of a reference device, taken at an angle. g) High magnification false color SEM image of the same reference device together with a circuit diagram schematic. The granularity visible on the picture is the surface of the gate metal.
}
\label{fig:Fig1}
\end{figure*}


\begin{figure*}
\includegraphics[width=\columnwidth]{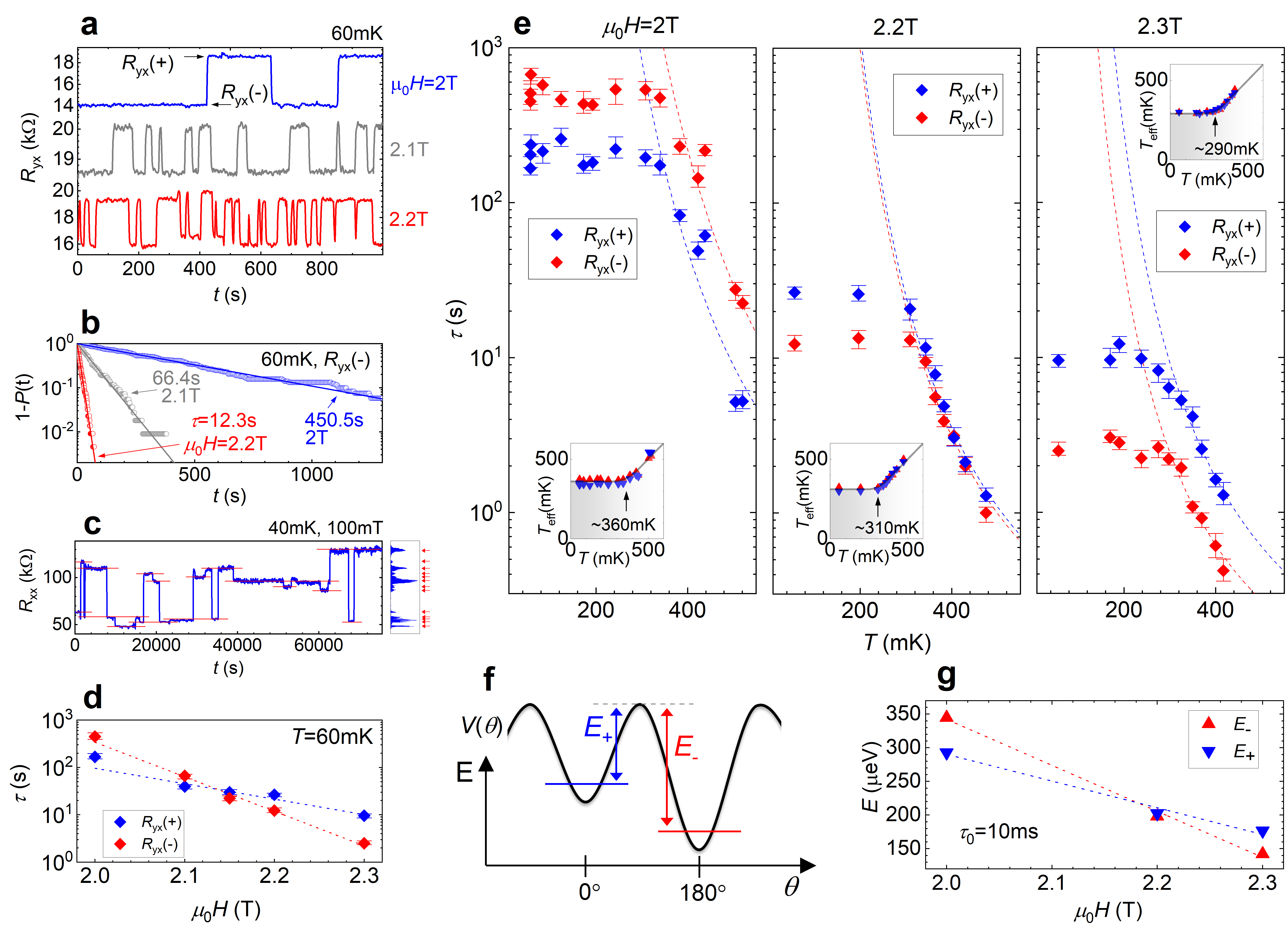}
\caption{
\textbf{Macroscopic quantum tunneling of magnetization.} a) Typical time dependent measurement of $R_{\mathrm{yx}}$ at various magnetic field and a temperature 60 mK, showing the telegraph noise in the signal. The "+" and "-" label the larger and smaller $R_{\mathrm{yx}}$ states, respectively. b) Probability of the state not being switched after a given time in the $R_{\mathrm{yx}}$(-) state, along with the fitting to an exponential decay function, for the three magnetic field values from (a). c) An $R_{\mathrm{xx}}$ measurement collected at a field of 100 mT, where we observe the largest number of discrete states (13) at base temperature. The horizontal red lines and the "bin counter" next to the figure help identify the number of distinct states. d) Evolution of the lifetime $\tau$ with magnetic field at 60 mK. e) Evolution of $\tau$ with temperature for three magnetic field values. The error bars represent a conservative range of values of $\tau$ that can describe the exponential decay. The insets show the effective temperature ($T_{\mathrm{eff}}$), saturating in the low temperature regime (see text for details). f) A simplified schematic of the anisotropic potential landscape for the magnetization, with energies defined: $E_{+}$ and $E_{-}$ are the thermal activation energy for each state. An angle $\theta$ of 0 or 180 degrees is along the easy axis of the ferromagnet (which is normal to plane of the sample). g) Evolution of the activation energies with the external magnetic field. The values correspond to the colored dashed lines in (e) following $\tau$=$\tau_0$exp[$E_{+/-}$/($k_{\mathrm{B}}T$)], with $\tau_0$=10 ms for all six curves.
}%
\label{fig:Fig2}%
\end{figure*}


\section[Introduction]{Introduction}

Since the birth of quantum mechanics in the early twentieth century, physicists have studied how the laws of quantum mechanics merge into those governing classical mechanics at macroscopic sizes. One of the more prominent consequences of quantum mechanics is the phenomenon of quantum tunneling between two separate eigenstates, an effect that has no direct analog in classical physics. This poses an intriguing question regarding the possibility of quantum tunneling effects in systems that can be regarded as macroscopic. In the past four decades, experimental techniques have started to reach the level of  sophistication needed to explore this issue. The topic has since been explored in studies of macroscopic quantum tunneling between different current states in Josephson junctions\cite{Caldeira1981,Voss1981,Devoret1985}, macroscopic quantum tunneling of the magnetization\cite{Chudnovsky1988,Awschalom1992,Coppinger1995,Friedman1996,Thomas1996,Sangregorio1997,Wernsdorfer1997,Barco1998}, and macroscopic quantum tunneling of magnetic domain walls\cite{Stamp1991,Brooke2001} in magnetic structures. The same period also saw the rise of topology in condensed matter\cite{Volkov1985,Hasan2010}, as topological insulators\cite{Kane2005,Hasan2010}, the first experimentally demonstrated topological state of matter\cite{Konig2007}, opened a path towards new quantum phenomena such as the quantum spin Hall effect\cite{Kane2005,Bernevig2006,Konig2007} in non-magnetic systems, and the quantum anomalous Hall effect\cite{Yu2010,Nomura2011,Chang2013} in ferromagnetic topological insulators. In this Article we investigate the quantum anomalous Hall state in a nanostructure with a mesa constriction as small as 160 nm, fabricated from a V-doped (Bi,Sb)$_2$Te$_3$ magnetic topological insulator layer. This size regime allows us to access and carefully investigate the dynamics of an individual magnetic domain in the material. The influence of temperature and magnetic field on the domain switching statistics shows that in the low temperature limit the switching is governed by macroscopic quantum tunneling of magnetization.

\section[Quantum anomalous Hall nanostructure device]{Quantum anomalous Hall nanostructure device}

Our magnetic topological insulator device is patterned from an 8.2 nm thick V$_{0.1}$(Bi$_{0.2}$Sb$_{0.8}$)$_{1.9}$Te$_3$ layer grown by molecular beam epitaxy (MBE) on a Si(111) substrate\cite{Winnerlein2017}. It is capped in-situ with a 10 nm thick Te layer, and the growth conditions are optimized for perfect anomalous Hall resistance quantization, as observed previously in our macroscopic-size devices\cite{Goetz2018,Grauer2015}. A high magnification scanning electron microscope (SEM) image of a reference device is shown in Fig. 1g. The six-terminal nanostructure has a width of about 160 nm and a length of about 530 nm (see blue arrows in Fig. 1g for how these dimensions are defined), and is patterned using a combination of electron beam lithography (for the mesa etched using Ar milling and AuGe ohmic contacts) and optical lithography (for the AlOx/HfOx/Au gate plus dielectric layer stack). While the sample is fitted with a top gate, for simplicity of the analysis the gate electrode is grounded for all experiments discussed in this Article (some measurements with an applied gate voltage can be found in Extended data Fig. 1). The influence of gate voltage on the physics will require further research and thus be the subject of future work.

\section[Results]{Results}

Fig. 1 presents basic magneto-transport data. The magnetic field direction is always perpendicular to plane of the sample. A low frequency AC voltage excitation with an amplitude of about 100 $\mu$V is applied between the source contact "S" and the grounded drain contact "D", while contacts 1 through 3 are used to probe the transverse (Hall)  $V_{\mathrm{yx}}$ and longitudinal $V_{\mathrm{xx}}$ voltages (a simplified circuit schematic is given in Fig. 1g). The current $I$ is obtained by measuring the voltage drop over a calibrated reference resistor (about 50 k$\Omega$) connected in series with the sample. The Hall resistance ($R_{\mathrm{yx}}$=$V_{\mathrm{yx}}$/$I$) of close to $\pm h$/$e^2$ visible in Fig. 1a shows that the device is in the quantum anomalous Hall regime. The small and fluctuating deviations from the perfectly quantized value are primarily a result of longitudinal voltage admixing. Indeed when the symmetric (with respect to the hysteretic $B$ field) component is subtracted from the measured $R_{\mathrm{yx}}$ signal, the accuracy of the quantization improves significantly (see Extended data Fig. 5). Since the larger Hall bars fabricated from this material exhibit robust quantum anomalous Hall quantization\cite{Goetz2018,Grauer2015}, the finite longitudinal resistance ($R_{\mathrm{xx}}$=$V_{\mathrm{xx}}$/$I$) visible in Fig. 1b may indicate the influence of sample size effects. For example, the edges of the device are close enough to each other to either directly couple due to the edge state wavefunction overlap\cite{Zhou2008}, or more likely due to a path of topological channels extending into the bulk by meandering through the magnetic domain structure. This of course also contributes to deviations of $R_{\mathrm{yx}}$ from an otherwise perfectly quantized value. In the end both measured $R_{\mathrm{xx}}$ and $R_{\mathrm{yx}}$ signals are likely a consequence of a complicated Landauer-B\"uttiker network of 1D channels formed around the magnetic domains profile within the material. An important point to note for the upcoming analysis is the clear difference in transport behavior observed between the 150 mK (Fig. 1c) and the 40 mK (Fig. 1a,b) measurements, implying that the lowest temperature experienced by the carriers in the sample is well below 150 mK in this experiment.

Fig. 2a highlights the key observation of this work. Here, the Hall resistance of the sample is recorded as a function of time, at three different values of magnetic field ($\mu_{\mathrm{0}}H$=2 T, 2.1 T, and 2.2 T), and at a temperature 60 mK. $R_{\mathrm{yx}}$ is manifestly not constant with time. This is in stark contrast to measurements in macroscopic (hundreds of microns wide) quantum anomalous Hall devices, where no instabilities are observed (except for during magnetic reversal at the coercive field of the ferromagnet)\cite{Liu2016}. The resistance instability has a two-level telegraph noise signature \cite{Coppinger1995,Wernsdorfer1997b,Coppinger1998}, suggestive of an individual magnetic domain switching between two magnetization states. For each of the measurements, the sample was magnetically prepared by first applying a field of 7.5 T, and ramping it back down to the specified value in the same direction, i.e. without crossing zero. We thus emphasize that the switching occurs spontaneously despite the magnetic field being still applied in the same direction as the sample magnetization, and thus being far away from the coercive field that would trigger a global magnetization reversal.

To properly analyze this telegraph noise, each measurement is allowed to run sufficiently long to record about 200 (or more) switching events. The duration of the plateaus between the switching events is then extracted, distinguishing between jumps from the "+" (larger $R_{\mathrm{yx}}$) or from the "-" (smaller $R_{\mathrm{yx}}$) state. From this data we can determine the probability $P$($t$) of the state having switched to the other state after having been in a given state for a time $t$. The probability of the state not having switched by time $t$ is then obviously 1-$P$($t$), which for a random process with a mean lifetime $\tau$ follows an exponential decay exp(-$t$/$\tau$)\cite{Wernsdorfer1997b,Wernsdorfer1997}. An example of this for the $R_{\mathrm{yx}}$(-) state is plotted as the symbols in Fig. 2b, along with an exponential decay fit, with the decay constant $\tau$ indicated in the figure giving the lifetime of the state under the given conditions. 

When time dependent measurements are performed at magnetic fields far away from the range focused on in Fig. 2 (which ranges from 2 T to 2.3 T), we observe magnetic switching between significantly more than 2 states, indicating that multiple magnetic domains are active. An example of this is shown in Fig. 2c for 100 mT, and similar behaviour is also observed when a gate voltage is applied; see Extended data Fig. 1. It thus is incidental that in the magnetic field range close to 2 T, only a single magnetic domain contributes to the instability. Given the high sensitivity of lifetime $\tau$ to the external magnetic field (as presented in Fig. 2d) it cannot be ruled out that other domains are also active, but that their contributions are either much too slow or much too fast, relative to the measurement time and data acquisition speed, to be perceived.

For an estimate of the size of the domain that controls the data in Fig. 2, we can turn to the data in Fig. 2c. The largest amount of distinct states observed in our device in any base temperature measurement occurs for a field of 100 mT, where 13 levels are seen (Fig. 2c). More measurements at different magnetic field values at base temperature can be found in Extended data Fig. 2, all showing less then 13 states. These 13 states imply at least 4 active domains (which could produce a maximum of 2$^4$=16 independent states). Given that the measured Hall signal never changes sign, one can assume that these 4 magnetic domains correspond to less than half of the device area. If the device area is taken as a 160 nm by 530 nm rectangle (blue arrows in Fig. 1g), this results in an upper bound on the magnetic domain size of about 115 nm in diameter, which, given the layer thickness of 8.2 nm corresponds to a volume of approx. 85 000 nm$^3$ (containing about 50 000 vanadium atoms). Interestingly, this is comparable to the size of crystal domains visible in atomic force microscope (AFM) measurements of the layer surface, hinting that the magnetic domains formed in this material system are commensurate to the crystal domains. An AFM scan from a reference uncapped (V,Bi,Sb)$_2$Te$_3$ layer grown under the same conditions can be found in Extended data Fig. 3. We stress that while this estimate provides an upper bound on the domain size, the real size cannot be significantly smaller then 115 nm as the number of states increases exponentially with reduced domain size. Assuming a domain size of about 80 nm in diameter, which is only a 30 $\%$ reduction, the number of independent states would increase to as many as 2$^8$=256, which already is extremely unlikely given only 13 observed levels in Fig. 2c.

It is natural to assume that the state with larger $R_{\mathrm{yx}}$ (labeled with the "+" index) represents a magnetic domain aligned in the same direction as the background magnetization, i. e. ferromagnetically coupled to it's surroundings (and thus adding to the total Hall signal), while the lower $R_{\mathrm{yx}}$ state (labeled "-") has the domain antiferromagnetically coupled to the surroundings. This interpretation is supported by the evolution of the lifetime $\tau$ of each state as a function of the external magnetic field, as analyzed in Fig. 2c. At a magnetic field of 2 T, the "-" state has the larger lifetime, and is therefore the energetically more stable state. As the magnetic field is increased, the lifetimes of the two states cross, and above about 2.15 T the "+" state becomes more stable, consistent with the external magnetic field favoring the ferromagnetic alignment of the domain, over the antiferromagnetic alignment. 

In Fig. 2e we turn to the temperature dependence of $\tau$ for each state, and for three magnetic fields. In the high temperature regime the lifetime follows an Arrhenius like thermal activation $\tau$=$\tau_{\mathrm{0}}$exp[$E_{+/-}$/($k_{\mathrm{B}}T$)] (colored dashed lines), where $E_{+/-}$ is the activation energy for each state, and $k_B$ the Boltzmann constant. This is consistent with the behavior expected from a thermally agitated single-domain magnetic particle\cite{Wernsdorfer1997b}. We find a good match for all 6 curves ("+" and "-" states for each magnetic field) with $\tau_{\mathrm{0}}$=10 ms. This is the value used for the dashed lines in the figure, but values ranging from 1 to 20 ms also describe the data reasonably well (see Extended data Fig. 4). The exact value of $\tau_{\mathrm{0}}$ quantitatively affects the obtained energies ($E_{+}$ and $E_{-}$), but the roughly linear dependence of these energies on magnetic field, shown in Fig. 2g, is robust to the uncertainty in $\tau_{\mathrm{0}}$. 

Note that the value of $\tau_{\mathrm{0}}$ observed here is significantly larger than a typical superparamagnetic attempt time, which is of the order of 1 ns\cite{Wernsdorfer1997b}. There are a number of possible contributing factors for this. First, the size of the magnetic domain is larger than that of typically investigated nanomagnets (about 115 nm in our case, about 25 nm in \cite{Wernsdorfer1997b}). Second, the magnetic domain is embedded in a system composed of many other nearby, and possibly interacting, magnetic domains, rather than being an isolated magnetic object. Finally the material at hand is a diluted magnetic system. All of those factors could influence the value of $\tau_{\mathrm{0}}$, and a full microscopic understanding of this difference may provide a valuable clue towards the elusive nature of magnetic interactions in this material. 

Much more interesting and fundamental however is the behavior in the low temperature regime. Here, the Arrhenius thermal activation picture breaks down, as the lifetime saturates, forming a large plateau in the temperature dependence of $\tau$. The dynamics are clearly no longer governed by thermal agitation, where the magnetization during the reversal process follows a classical trajectory over the anisotropy induced energy barrier. Rather, the quantum limit has been reached, and the switching rate depends solely on the tunneling probability through the potential barrier\cite{Chudnovsky1988}, a trajectory forbidden in classical physics. Indeed, such a low temperature saturation of magnetic dynamics, followed by a high temperature thermally activated regime, was previously attributed to quantum tunneling in a number of magnetic systems\cite{Coppinger1995,Sangregorio1997,Wernsdorfer1997,Coppinger1998,Brooke2001}. In order to better identify the transition temperature, the data of Fig. 2e can be used to generate it's insets. Here the $x$ axis gives the temperature for each measurement, whereas the $y$ axis gives an effective temperature ($T_{\mathrm{eff}}$) which would be associated with the given lifetime within the thermal activation model. Plotted in this way, a clear inflection between the two regimes is visible, as marked by the arrows in the insets.

As visible in Fig. 2g, the thermal activation energies (and therefore the potential barrier height for each state), as extracted from the high temperature regime, decrease significantly as magnetic field is increased (from about 300 $\mu$eV to 150 $\mu$eV when the magnetic field is changed from 2 T to 2.3 T). A quantum mechanical wavefunction penetrating into a potential barrier, to a good approximation, decays exponentially, with a decay constant strongly dependent on the barrier height. Therefore the tunneling probability is expected to significantly increase when the barrier height is moderately reduced. This explains the roughly exponential decrease of $\tau$ for each state with increasing magnetic field that is shown in Fig. 2c for 60 mK, a temperature deep in the quantum tunneling regime. In addition we observe a decrease in the onset point of saturation of $T_{\mathrm{eff}}$ with increased magnetic field. The insets in Fig. 2e show that the $T_{\mathrm{eff}}$ plateau develops at a temperature of about 360 mK at magnetic field of 2 T, at about 310 mK at 2.2 T , and at about 290 mK at 2.3 T. This is likely a result of a complicated balance between the tunneling rate and the thermal agitation rate, as both can in principle evolve differently with the potential barrier height. Nevertheless the fact that the observed transition temperature reacts to the magnetic field, further demonstrates that the observed low temperature saturation of dynamics is governed by the magnetic properties of the device.

\section[Discussion]{Discussion}

Over the past 25 years or so, investigations into quantum tunneling of magnetization have primarily focused on tunneling in large ensembles of magnetic molecules\cite{Friedman1996,Thomas1996,Sangregorio1997,Henderson2009}. We stress that while quantum tunneling in such individual molecules is frequently referred to as "macroscopic", it is fundamentally a collective response of an ensemble of identical microscopic objects. Each molecule has a size of order 1 nm, and a spin resulting from only the handful of magnetic atoms within the molecule. This leaves open room for debate as to how truly "macrosopic" such tunneling is. Only a few works have reported on quantum tunneling of magnetization in somewhat larger magnetic structures. This includes molecules of antiferromagnetic ferritin where the magnetic core, which is comprised of up to 4500 Fe atoms, has a diameter of only about 7 nm\cite{Barco1998}, as well as lithographically patterned SrRuO$_3$ nanostructures where the tunneling volume was found to be close to 100 nm$^3$\cite{Sinwani2012} (i.e. a few nanometers in diameter). The only reported evidence for quantum tunneling of magnetization observed in an individual magnetic nanocrystal (ferrimagnetic barrium ferrite) can be found in Ref. \cite{Wernsdorfer1997}. There, though not as clearly pronounced as in our results, the Authors observed a clear onset of saturation of the magnetic dynamics at low temperature, corresponding to macroscopic quantum tunneling. The system size there is about 10-20 nm in diameter.

In contrast, we observe quantum tunneling of a ferromagnetic macrospin originating from a dillute ferromagnetic domain containing as many as 50 000 magnetic atoms (out of 2 500 000 atoms in total), spread over a clearly macroscopic distance of 115 nm (and a volume of 85 000 nm$^3$), making it, by volume, some 2 orders of magnitude larger than any previously studied macrospin. Our results thus not only provide the most robust evidence of quantum tunneling of a truly macroscopic magnetic state reported to date, but also provide first realization of this effect in a topological state of matter. This in turn opens a path towards investigating possible intriguing connections between macroscopic quantum phenomena and topology.

\section[Summary]{Summary}

We have fabricated a (V,Bi,Sb)$_2$Te$_3$ topological insulator nanostructure in the quantum anomalous Hall regime. The size of the device is small enough that dynamics of an individual ferromagnetic domain can be resolved, with an expected (based on the magnetic domain volume estimate of approx. 85 000 nm$^3$) collective magnetic moment originating from about 50 000 vanadium atoms spread over a macroscopic distance of 115 nm. Careful analysis of the influence of magnetic field and temperature on the telegraph noise originating from the domain magnetization fluctuations, provides strong evidence of quantum tunneling of a truly macroscopic magnetization state, and that in a topological insulator.

\bigskip

\section[Acknowledgents]{Acknowledgents}
We gratefully acknowledge the financial support of the Free State of Bavaria (the Institute for Topological Insulators), Deutsche Forschungsgemeinschaft (SFB 1170, 258499086), W\"urzburg-Dresden Cluster of Excellence on Complexity and Topology in Quantum Matter (EXC 2147, 39085490), and the European Commission under the H2020 FETPROACT Grant TOCHA (824140).

\section[Author contributions]{Author contributions}
K. M. F. designed and patterned the measured device, performed the transport experiments, and the telegraph noise analysis. N. L. grew the sample. P. M. contributed to nanolithography development and fabricated the reference device. S. S. contributed to growth optimization. K. B., C. G., and L. W. M. supervised the work. All authors contributed to the analysis and interpretation of the results, and the writing of the manuscript.

\setcounter{figure}{0} 
\renewcommand{\figurename}{Extended data Fig.}

\begin{figure*}
\includegraphics[width=\columnwidth]{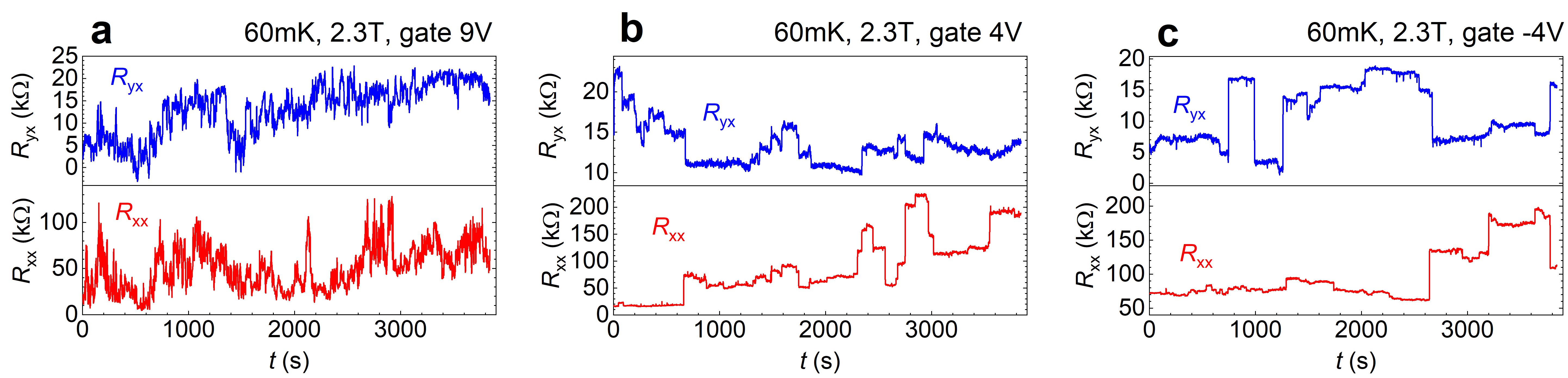}
\caption{
Time dependent measurements of Hall and longitudinal resistance collected with different applied gate voltages, at a constant magnetic field of 2.3 T and temperature 60 mK. The applied gate voltage is: a) 9V, b) 4V, and c) -4V.
}
\label{fig:EDFig1}
\end{figure*}

\begin{figure*}
\includegraphics[width=\columnwidth]{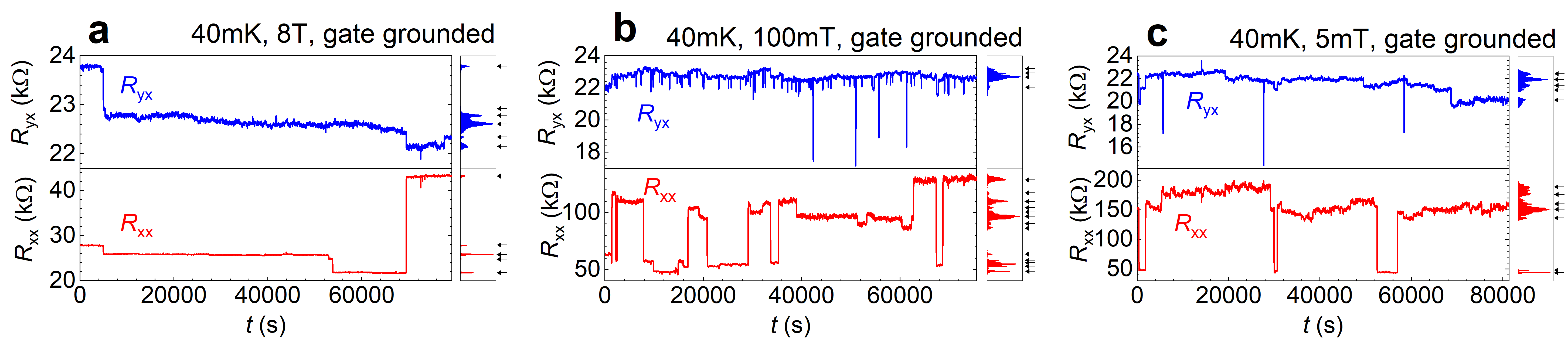}
\caption{
Time dependent measurements of Hall and longitudinal resistance collected with a grounded gate, at different magnetic field values, and a constant temperature of 40 mK. The external magnetic field is: a) 8T, b) 100 mT, and c) 5 mT. A "bin counter" is plotted next to each figure, to identify the number of distinct states. The largest number of distinct states, 13, is visible in (b) $R_{\mathrm{xx}}$.
}
\label{fig:EDFig2}
\end{figure*}

\begin{figure*}
\includegraphics[width=5in]{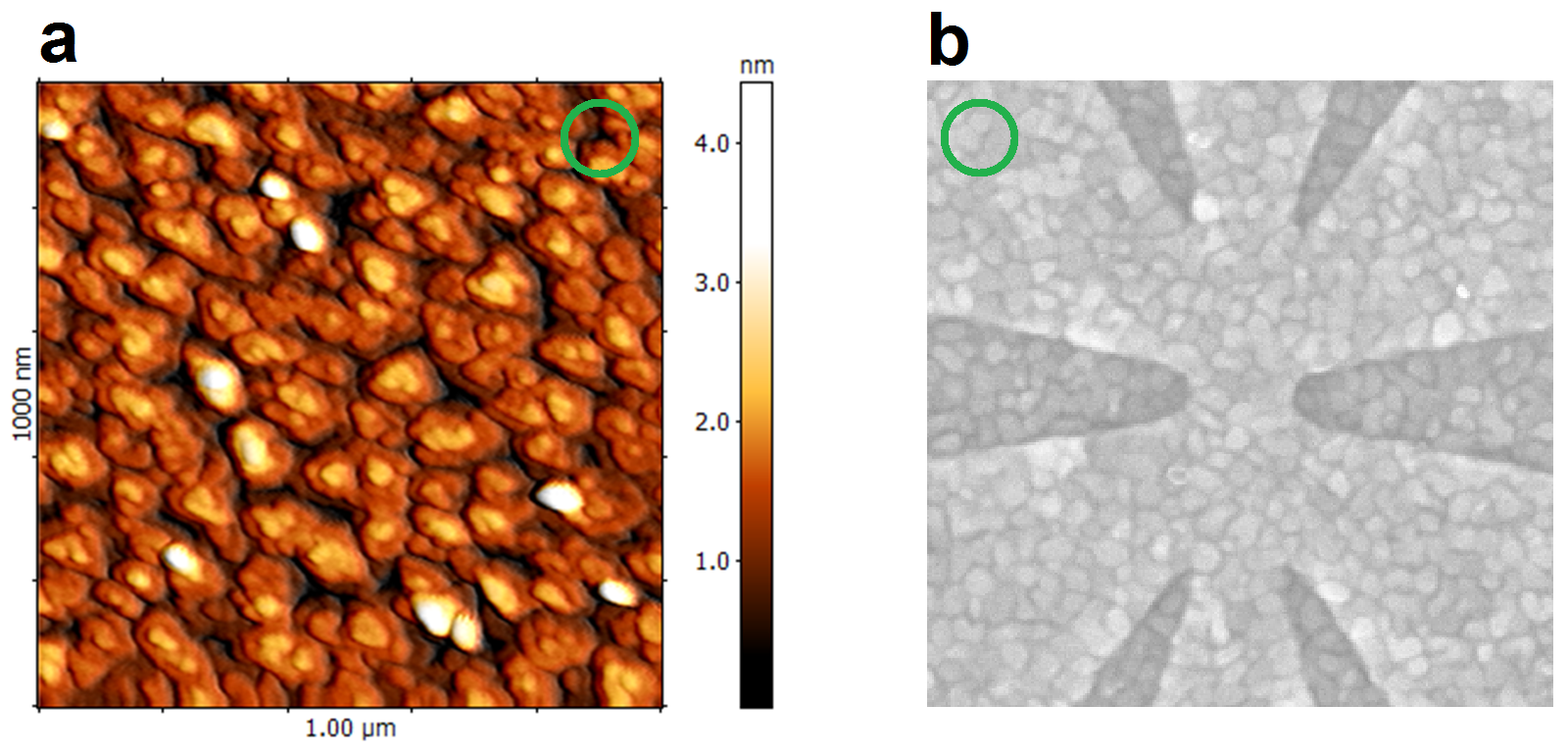}
\caption{
Comparison between an estimated magnetic domain size and the surface topography of the film. a) Atomic field microscope (AFM) image of the surface topography of a reference (V,Bi,Sb)$_2$Te$_3$ layer grown under the same conditions (but without the protective Te cap) as the layer from which the investigated nanostructure was patterned. b) Scanning electron microscope (SEM) image of a reference device patterned using the same lithographic process as the device investigated in the Article. Both images are 1 $\mu$m by 1 $\mu$m. The green circles represent the estimated upper bound for the magnetic domain size of diameter 115 nm. A granularity visible in picture (b) is from the surface of the gate metal. 
}
\label{fig:EDFig3}
\end{figure*}

\begin{figure*}
\includegraphics[width=\columnwidth]{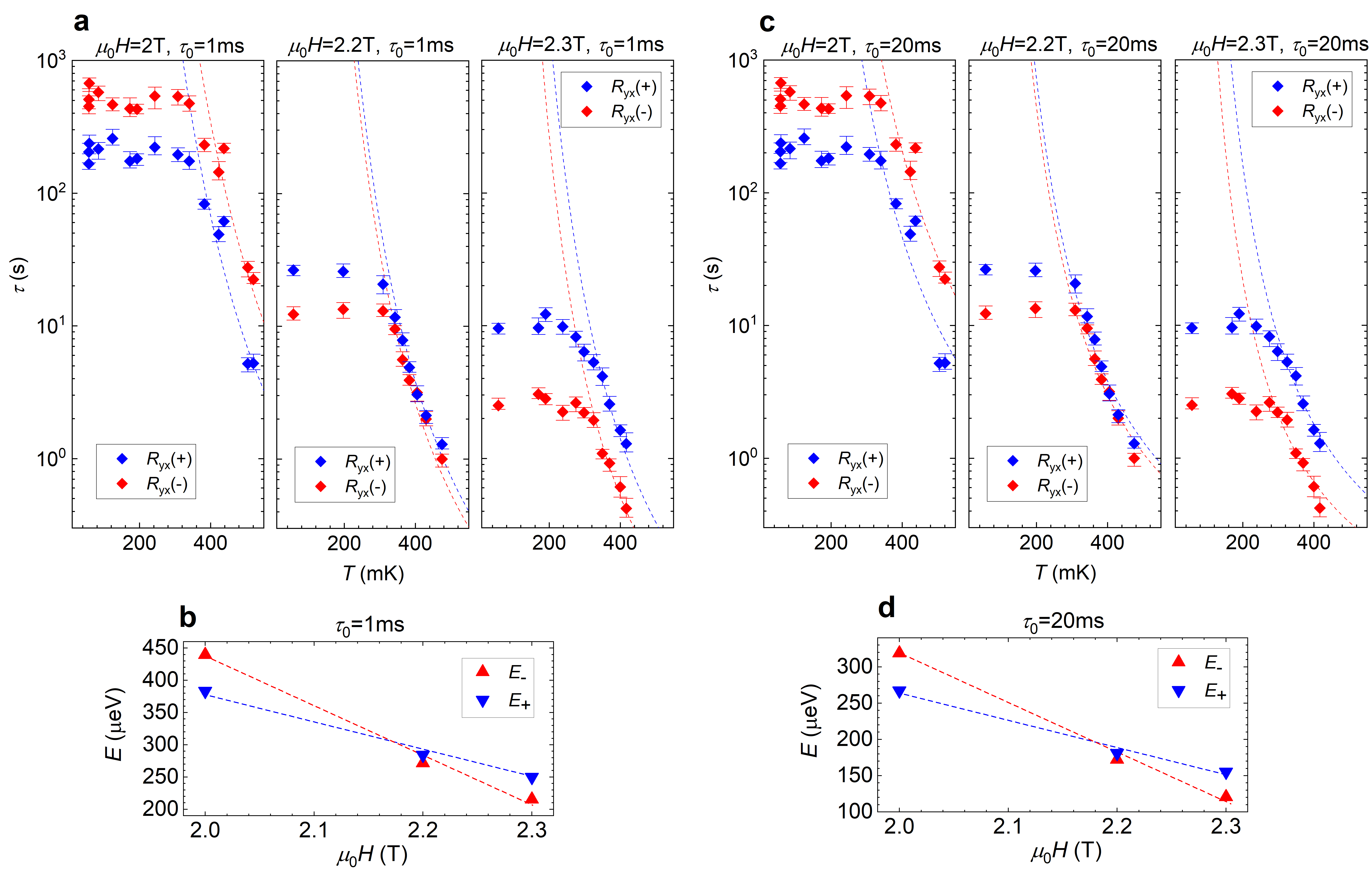}
\caption{
Thermal activation analysis for different values of parameter $\tau_{\mathrm{0}}$. a) Evolution of $\tau$ with temperature for three different magnetic field values with high temperature activation dashed lines for $\tau_{\mathrm{0}}$=1 ms, and b) corresponding thermal activation energies $E_{+}$ and $E_{-}$. c) Evolution of $\tau$ with temperature for three different magnetic field values with high temperature activation dashed lines for $\tau_{\mathrm{0}}$=20 ms, and d) corresponding thermal activation energies $E_{+}$ and $E_{-}$. The dashed lines in (a) and (c) follow $\tau$=$\tau_{\mathrm{0}}$exp[$E_{+/-}$/($k_{\mathrm{B}}T$)]. The dashed lines in (b) and (d) are guide to the eyes.
}
\label{fig:EDFig4}
\end{figure*}

\begin{figure*}
\includegraphics[width=\columnwidth]{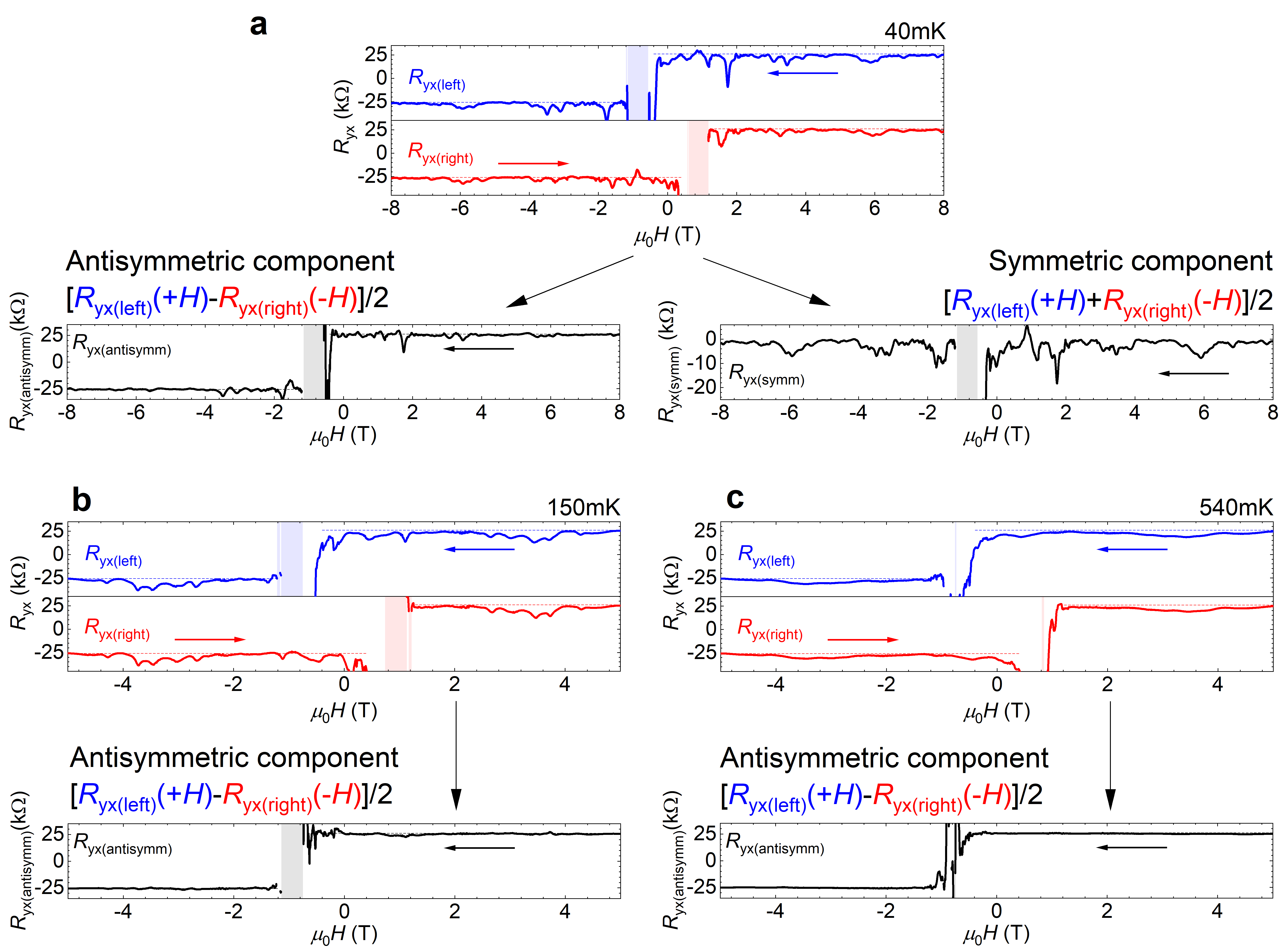}
\caption{
a) The top panel repeats the $R_{\mathrm{yx}}$ data from Fig. 1a (collected at 40 mK), which in the bottom part of the figure is split into antisymmetric and symmetric components with respect to the hysteretic $B$ field. In order to account for hysteresis, the components are extracted by calculating the average (or difference) between the sweep towards the left for a given $H$ field, with the sweep to the right for the negative of that field. b) The same for the antisymmetric component for the $R_{\mathrm{yx}}$ data collected at 150 mK (reproduced from Fig. 1c), and c) for the $R_{\mathrm{yx}}$ data collected at 540 mK (reproduced from Fig. 1d). The arrows in the figures indicate the sweep direction. The shaded regions represent a regime around the global magnetization reversal where the sample is insulating. The horizontal dashed lines represent resistance values expected for a perfect quantum anomalous Hall effect. 
}
\label{fig:EDFig5}
\end{figure*}

\bigskip

\end{document}